\documentclass[twocolumn]{aastex62}
\usepackage{amsmath}
\usepackage[normalem]{ulem}

\shorttitle{On the Circular Polarisation from Repeating FRBs}
\shortauthors{Dai et al.}

\begin{document}

\title{On the Circular Polarisation of Repeating Fast Radio Bursts}

\correspondingauthor{Shi Dai}
\email{s.dai2@westernsydney.edu.au}

\author[0000-0002-9618-2499]{Shi Dai}
\affiliation{Western Sydney University, Locked Bag 1797, Penrith South DC, NSW 2751, Australia}
\affiliation{CSIRO Space and Astronomy, Australia Telescope National Facility, Epping, NSW 1710, Australia}

\author[0000-0002-9815-5873]{Jiguang Lu}
\affiliation{National Astronomical Observatories, Chinese Academy of Sciences, Beijing 100101, People's Republic of China}

\author[0000-0002-9815-5873]{Chen Wang}
\affiliation{National Astronomical Observatories, Chinese Academy of Sciences, Beijing 100101, People's Republic of China}

\author[0000-0001-9036-8543]{Wei-Yang Wang}
\affiliation{School of Physics and State Key Laboratory of Nuclear Physics and Technology, Peking University, Beijing 100871, People's Republic of China}
\affiliation{National Astronomical Observatories, Chinese Academy of Sciences, Beijing 100101, People's Republic of China}

\author[0000-0002-9042-3044]{Renxin Xu}
\affiliation{School of Physics and State Key Laboratory of Nuclear Physics and Technology, Peking University, Beijing 100871, People's Republic of China}
\affiliation{Kavli Institute for Astronomy and Astrophysics, Peking University, Beijing 100871, People's Republic of China}
\affiliation{Department of Astronomy, School of Physics, Peking University, Beijing 100871, People's Republic of China}

\author[0000-0001-6374-8313]{Yuanpei Yang}
\affiliation{South-Western Institute for Astronomy Research, Yunnan University, Kunming 650500, Yunnan, People's Republic of China}

\author[0000-0003-2366-219X]{Songbo Zhang}
\affiliation{Purple Mountain Observatory, Chinese Academy of Sciences, Nanjing 210008, People's Republic of China}

\author[0000-0003-1502-100X]{George Hobbs}
\affiliation{CSIRO Space and Astronomy, Australia Telescope National Facility, Epping, NSW 1710, Australia}

\author[0000-0003-3010-7661]{Di Li}
\affiliation{National Astronomical Observatories, Chinese Academy of Sciences, Beijing 100101, People's Republic of China}
\affiliation{University of Chinese Academy of Sciences, Beijing 100049, People's Republic of China}
\affiliation{FAST Collaboration, CAS Key Laboratory of FAST, NAOC, Chinese Academy of Sciences, Beijing 100101, People's Republic of China}

\author[0000-0002-4300-121X]{Rui Luo}
\affiliation{CSIRO Space and Astronomy, Australia Telescope National Facility, Epping, NSW 1710, Australia}

\author[0000-0002-4990-9288]{Miroslav Filipovic}
\affiliation{Western Sydney University, Locked Bag 1797, Penrith South DC, NSW 2751, Australia}

\author[0000-0002-6465-0091]{Jinchen Jiang}
\affiliation{Department of Astronomy, School of Physics, Peking University, Beijing 100871, People's Republic of China}

\begin{abstract} 
Fast spinning (e.g., sub-second) neutron star with ultra-strong magnetic fields (or so-called magnetar) is one of the promising origins of repeating fast radio bursts (FRBs). Here we discuss circularly polarised emissions produced by propagation effects in the magnetosphere of fast spinning magnetars.
We argue that the polarisation-limiting region is well beyond the light cylinder, suggesting that wave mode coupling effects are unlikely to produce strong circular polarisation for fast spinning magnetars. Cyclotron absorption could be significant if the secondary plasma density is high. However, high degrees of circular polarisation can only be produced with large asymmetries in electrons and positrons. 
We draw attention to the non-detection of circular polarisation in current observations of known repeating FRBs. We suggest that the circular polarisation of FRBs could provide key information on their origins and help distinguish different radiation mechanisms.

\end{abstract}


\section{Introduction}
Fast Radio Bursts (FRBs) are bright millisecond-duration radio transients first discovered by \citet{lbm+07}. Their cosmological origin and energetic nature make them ideal tools to probe a range of astrophysical and fundamental physics~\citep[e.g.,][]{rsb+16,pmm+19,mpm+20}. Our knowledge of the progenitors of FRBs and the radiation mechanism is limited and whether repeating and non-repeating FRBs share the same origin is still an open question~\citep[see][for a review]{zha20}. Recently, \citet{pgk+21} presented a synthesis of morphology of 18 repeating FRBs and 474 non-repeating FRBs. They showed that bursts from repeating FRBs, on average, have larger widths and are narrower in bandwidth. Comparisons of FRB luminosity functions with predictions based on event rate densities of various models also show that it is hard to explain both repeating and non-repeating FRBs with the same origin~\citep[e.g.,][]{lml+20}. Precise localisation of repeating and non-repeating FRBs~\citep[e.g.,][]{clw+17,bdp+19,rcd+19,mnh+20,kmn+21,rll+21,fdl+21} showed that properties of FRB host galaxies and local environments are diverse~\citep[e.g.,][]{tbc+17,bsp+20,lz20,hps+20}.

Most theories of active repeating FRBs involve neutron stars as their central engine. They can be young/normal pulsars~\citep[e.g.,][]{cw16,csp16,yz18,wang+19,yzz+20,wxc20}, or pulsars with ultra-strong magnetic fields/magnetars~\citep[e.g.,][]{kat16,dwu+16,bel17}, or normal pulsars with external~\citep[e.g.,][]{zha17,mzv20} or internal~\citep[e.g.,][]{wly+18} interactions. Pulsars with short spin periods and ultra-strong magnetic fields have drawn much attention since they are likely young and store a large amount of toroidal magnetic energy inside the star, which could explain their active bursting activities~\citep[e.g.,][]{mbs+20,lkz20}. It has been proposed that rare, extreme explosions such as long gamma-ray bursts~\citep[LGRBs,][]{zm01,mgt+11}, superluminous supernovae~\citep[SLSNe; e.g.,][]{mkm16,mbm17}, or NS mergers~\citep[e.g.,][]{mbm19,jwl+20,wwy+20} could produce such fast spinning pulsars with ultra-strong magnetic fields. Theories involve magnetars are supported by the recent detection of a FRB (FRB 200428) from the well-known Galactic magnetar SGR 1935$+$2154~\citep{abb+20,brb+20}. Magnetars as the origin of repeating FRBs were likely born with a faster initial spin and more toroidal magnetic energy inside the star, and are therefore more active compared with Galactic magnetars~\citep[e.g.,][]{lzw+20,mbs+20,lkz20,kat20}.

In addition to studies of temporal and frequency structures of FRBs, radio polarisation properties of FRBs have been presented in a number of papers~\citep[e.g.,][]{pbb+15,mls+15,ckv+18,msh+18,fab+20,dds+20,luo+20,nhk+21}. These observations and studies of radiation mechanism~\citep[e.g.,][]{cw16,lkn19} have suggested that polarisation properties might provide crucial information on the emission mechanism of FRBs. Propagation effects in the potentially dense and relativistic magneto-ionic environments of FRBs, such as Faraday conversion, have also been investigated. \citet{vr19} used the non-detection of Faraday conversion in FRB~121102 to constrain FRB progenitor models and the magnetic field in the cold confining medium. \citet{gl19} provided qualitative predictions for the circular polarisation of FRBs produced by Faraday conversion and their narrow-band properties at low frequencies. 

In this paper, we investigate the origin of circular polarisation of fast spinning magnetars and discuss its implication on repeating FRBs. We focus on wave mode coupling and cyclotron absorption effects and argue that it is challenging to produce high degrees of circular polarisation for fast spinning magnetars through propagation effects. Future studies (both observational and theoretical) of circular polarisation of a large sample of repeating FRBs might reveal key information on their origins and radiation mechanisms. In Section~\ref{sec:obs}, we will discuss observational properties of circular polarisation of repeating and non-repeating FRBs. In Sections~\ref{sec:prop} and \ref{sec:magnetar}, propagation effects will be discussed for the fast spinning magnetar model. Discussions and conclusions will be given in Section~\ref{sec:con}.

\section{Circular polarisation of radio pulsars and FRBs}
\label{sec:obs}

\begin{figure}
\centering
\includegraphics[width=0.5\textwidth]{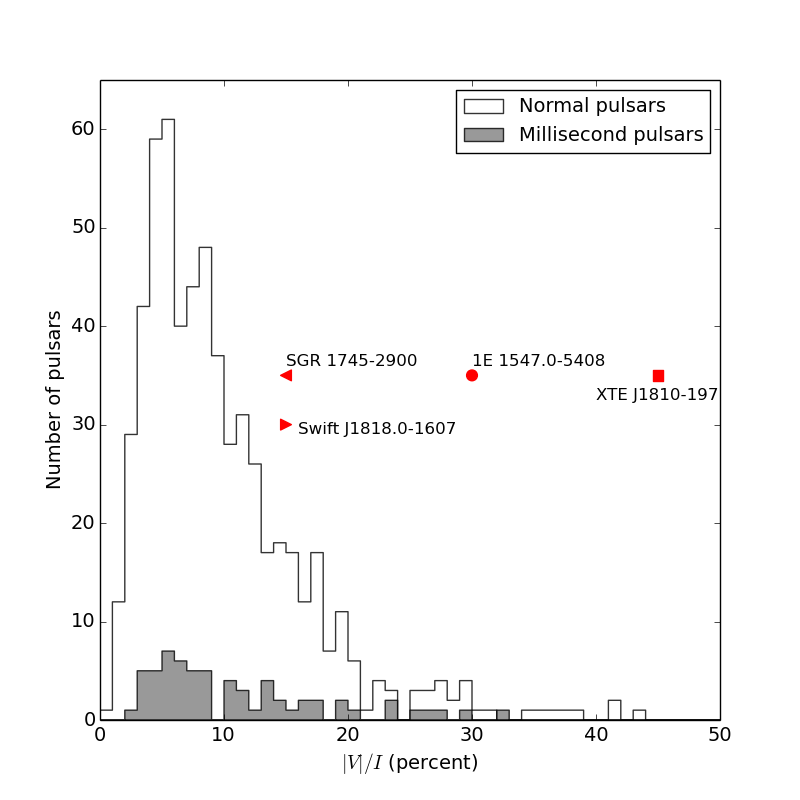}
\caption{Histograms of fractional circular polarisation ($|V|/I$) of normal pulsars (black line) and millisecond pulsars (filled region) at $\sim1.4$\,GHz. Red points represent the fractional circular polarisation of four radio-loud magnetars. We note that the circular polarisation of magnetars often shows large variations, and values shown here represent the high end of their fractional circular polarisation.}
\label{fig:hist}
\end{figure}

Compared with other radio sources (e.g., active galactic nucleus), radio pulsars as a population show significantly stronger circular polarisation. Fig.~\ref{fig:hist} shows histograms of fractional circular polarisation ($|V|/I$) of 600 normal pulsars~\citep{jk18} and 63 millisecond pulsars at $\sim1.4$\,GHz~\citep{dhm+15,gmd+18,wmg+21}. We can see that the majority of this sample of pulsars show a fractional circular polarisation of $\sim5$ to 20 percent. More relevant to our discussion here is radio-loud magnetars. In Fig.~\ref{fig:hist}, red points show published fractional circular polarisation of four radio-loud magnetars\footnote{We note that high fractional circular polarisation have also been observed in Category III and IV profiles of PSR J1622$-$4950~\citep{lbb+12}, but exact numbers were not published in the literature. For XTE J1810$-$197 and Swift J1818.0$-$1607, we used fractional circular polarisation averaged across their wideband observations.}~\citep[e.g.,][]{crj+08,efk+13,dlb+19,lsj+20}. Although only five radio-loud magnetars are known so far, circular polarisation stronger than the average of normal pulsars have been detected from all of them. After their outbursts, rapid changes in the strength and sign of circular polarisation have been observed along with variations in linear polarisation and position angles~\citep[e.g.,][]{crj+07,lbb+12,djw+18,dlb+19}. While several mechanisms have been proposed to explain observed circular polarisation from normal radio pulsars (see discussions in Section~\ref{sec:prop}), the case of magnetar has not been explored in detail in the literature.

Polarised radio emission of FRBs shares many similarities with those of radio pulsars and magnetars. Variations in both linear and circular polarisation across the pulse have been observed in non-repeating FRBs~\citep[e.g.,][]{cms+20,dds+20}, and polarised emission of several cases are dominated by linear polarisation~\citep[e.g.,][]{ckv+18,cms+20}. Strong circular polarisation have been observed in a good fraction of non-repeating FRBs~\citep[e.g.,][]{pbb+15,mls+15,pbk+17,ckv+18,cms+20,dds+20}. In a case study of FRB 181112, \citet{cms+20} suggested that the variation in the circular polarisation across the profile provided evidence of radiation propagating through a relativistic plasma in the source region. They pointed out that such propagation effects could be common for non-repeating FRBs.

Some repeating FRBs also show high degrees of linear polarisation. Diverse polarisation angle swings observed in FRB 180301 is particularly reminiscent of polarisation properties of magnetars~\citep{luo+20}. However, despite deep multi-wavelength radio observations of FRBs 121102~\citep{ssh+16,msh+18,hms+21}, 171019~\citep{kso+19}, 180916.J0158$+$65~\citep{cab+20,nhk+21}, 180301~\citep{luo+20}, 20190711A~\citep{ksf+21}, and CHIME repeating FRBs~\citep{chime_r19,fab+20}, no significant circular polarisation has been detected. Although only two very active repeating FRBs (121102 and 180301) are known so far, the non-detection of circular polarisation from any of their bursts is puzzling for models involving fast spinning magnetars. It is intriguing to understand why fast spinning magnetars, assuming that they are the origin of repeating FRBs, are seemingly lack of strong circular polarisation while strong and highly variable circular polarisation has been observed in all Galactic radio-loud magnetars.

\section{Circular polarisation produced by propagation effects}
\label{sec:prop} 

The origin of circular polarisation from radio pulsars is a long-standing problem. Although many observational properties of pulsar circular polarisation are not fully understood, it is generally believed that propagation effects in magnetospheric plasma play an essential role and could naturally produce circular polarisation~\citep[e.g.,][]{wlh10,bp12}. Propagation effects that have been widely discussed in the literature include
\begin{itemize}
    \item Wave mode coupling~\citep[e.g.,][]{cr79,sti82}. In the vicinity of the emission origin, linearly polarised ordinary and extraordinary modes of normal waves propagate independently. In rotating magnetospheres with strong magnetic fields and relativistic plasma, the polarisation plane of normal waves is tilted to the ambient magnetic line planes, which produces observable circular polarisation. As the emission propagates, the plasma density decreases and eventually stops influencing the waves. This is where the polarisation becomes fixed, and is characterised by the so-called polarisation–limiting radius, $r_{\rm pl}$. Studies of wave mode coupling and polarisation transfer in relativistic plasma have been published in several papers~\citep[e.g.,][]{bar86}.
    \item Cyclotron absorption. Cyclotron absorption of radio emission within pulsar magnetospheres has been studied by \citet{bs76}, \citet{lm01} and \citet{flm03}. Cyclotron resonance occurs when the wave frequency in the electron/positron rest frame is close to the cyclotron frequency. When there is an asymmetry between electrons and positrons, intensities of left- and right-handed waves are different after the cyclotron absorption, which results in circular polarisation. In the inner magnetosphere, the Doppler shifted cyclotron frequency is much higher than the wave frequency. However, as waves propagate, the cyclotron frequency decreases and cyclotron absorption becomes significant when the cyclotron frequency is equal to the wave frequency. Such a condition occurs at the resonance radius, $r_{\rm cyc}$.  
\end{itemize}
Another effect that has been investigated in the literature is circularisation of natural waves~\citep[e.g.,][]{vlk98,pl00,wlh10,bp12}. Circular polarisation is generated as natural waves propagating through pulsar magnetospheres along curved field lines. However, as shown by \citet{vlk98}, for a strong magnetic field, circularisation only happens when the wave vector is nearly aligned with the magnetic field~\citep[also see discussions in][]{wlh10}. For a relativistic plasma, the intersecting angle between the propagating beam and the magnetic field is Lorentz-transformed (i.e. the angle increases), which suggests that natural waves are more likely to be highly linearly polarised~\citep{vlk98}.

In this paper, we focus on wave mode coupling and cyclotron absorption effects for the special case of fast spinning magnetars. For normal radio pulsars (with a spin period of $\sim1$\,s and a dipole magnetic field of $\sim10^{12}$\,G), \citet{wlh10} carried out detailed studies of various propagation effects and concluded that the observed circular polarisation is determined by the wave mode coupling while cyclotron absorption only changes the total intensity. \citet{wlh10} also showed that, for typical pulsar magnetosphere parameters, such as a Lorentz factor $\gamma$ of $\sim100$ and a plasma density of $\eta\equiv N/N_{\rm GJ}\sim100$, characteristic radii of various propagation effects follow
\begin{equation}
\label{eq:r}
    r_{\rm pl} < r_{\rm cyc} \lesssim r_{\rm lc},
\end{equation}
where $r_{\rm lc}=c/\Omega$ is the radius of the light cylinder and $\Omega$ is the angular spin frequency. In the following section, we will show that $r_{\rm pl}$ and $r_{\rm cyc}$ are most likely to be much larger than $r_{\rm lc}$ for fast spinning magnetars, which will have important implications on the degree of circular polarisation expected in the observed radio emission.

\section{Propagation effects for fast spinning magnetars}
\label{sec:magnetar}

Fast spinning (e.g., sub-second) neutron stars with ultra-strong magnetic fields (or so-called magnetars) have attracted much attention in order to understand the origin of FRBs, particularly for repeating FRBs~\citep[e.g.,][]{lkz20}. While ultra-strong magnetic fields and short spin periods can help explain repeating burst activities, they imply that pulsar magnetospheres are small (i.e. small $r_{\rm lc}$) and regions where wave mode coupling and cyclotron absorption occur are far away from the star (i.e. large $r_{\rm pl}$ and $r_{\rm cyc}$). 

Here we adopted solutions of the magnetic field structure and evaluations of $r_{\rm pl}$ and $r_{\rm cyc}$ given by \citet{bar86}. Outside the light cylinder, \citet{bar86} assumed that plasma flows in a nearly radial direction (the plasma density follows $\eta\propto r^{-2}$) and the magnetic field follows $B_{\rm r}\approx B_{\rm r}(r_{\rm lc})(r_{\rm lc}/r)^2$. While the structure of magnetic field and properties of plasma outside the light cylinder of fast spinning magnetars could be much more complex than this, \citet{bar86} provided a self-consistent solution and allows us to carry out an order-of-magnitude estimate. It is worth noting that most previous studies of the origin of circular polarisation focused on propagation effects within the light cylinder of normal pulsars, and the case of fast spinning magnetars has not been discussed before.

\begin{figure}
\centering
\includegraphics[width=0.5\textwidth]{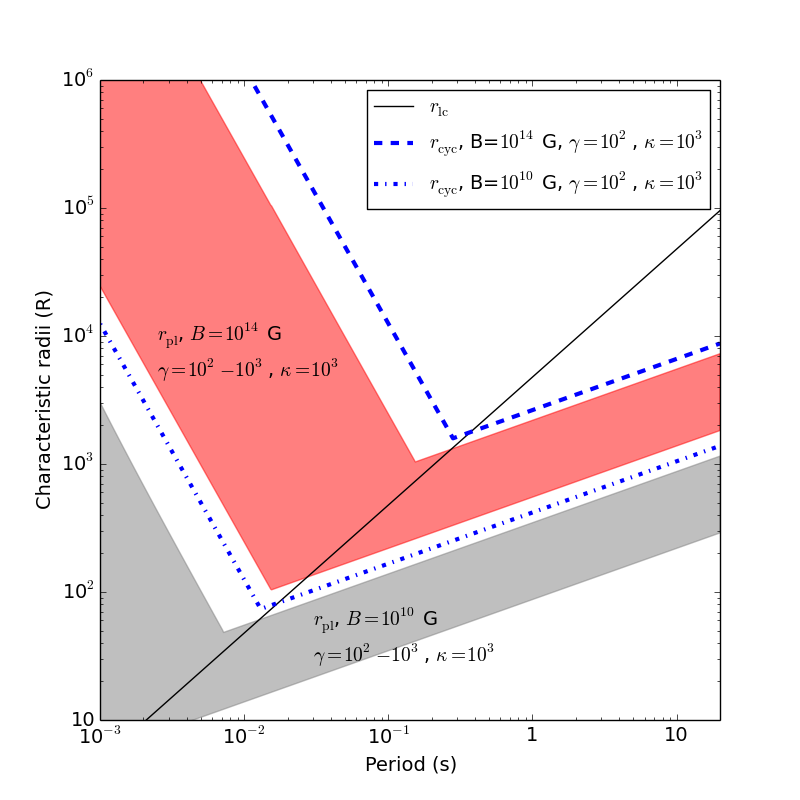}
\caption{Polarisation limiting radius, $r_{\rm pl}$, resonance radius, $r_{\rm cyc}$ and light cylinder radius, $r_{\rm lc}$ as a function of pulsar spin period. $r_{\rm pl}$ and $r_{\rm cyc}$ are estimated for magnetic field strength of $10^{10}$ and $10^{14}$\,G. Shaded regions show $r_{\rm pl}$ with $\gamma$ in the range of $10^2$ to $10^3$.}
\label{fig:rad}
\end{figure}

Following \citet{bar86}, we estimate the cyclotron absorption radius $r_{\rm cyc}$ as
\begin{equation}
    \frac{r_{\rm cyc}}{R}\approx\begin{cases}
   1050 B_{12}^{1/5}\nu_{9}^{-1/5}\gamma_{2}^{-1/5}P^{2/5}     &  \mbox{for}\  r\lesssim r_{\rm lc}\\
    1.25 B_{12}\nu_{9}^{-1}\gamma_{2}^{-1}P^{-2}     &  \mbox{for}\  r>r_{\rm lc}, \\
   \end{cases} 
\end{equation}
where $R=10^6$\,cm is the radius of the neutron star, $B_{12}$ is the strength of magnetic field in units of $10^{12}$\,G, $\gamma_{2}=\gamma/100$, $\nu_{9}$ is the radio frequency in $10^9$\,Hz and $P$ is the spin period in seconds. For a typical Lorentz factor of $\gamma\approx100-1000$ and a plasma density of $\eta\approx100-1000$, the cyclotron absorption radius $r_{\rm cyc}$ is always larger than the polarisation-limiting radius $r_{\rm pl}$~\citep{bar86,wlh10}. Under this scenario, the polarisation-limiting radius $r_{\rm pl}$ can be estimated as
\begin{equation}
\label{eq:pl}
        \frac{r_{\rm pl}}{R}\approx\begin{cases}
   873\kappa_{3}^{1/5}B_{12}^{1/5}\nu_{9}^{-1/5}\gamma_{2}^{-3/5}P^{2/5}     &  \mbox{for}\  r\lesssim r_{\rm lc}\\
    0.246\kappa_{3} B_{12}\nu_{9}^{-1}\gamma_{2}^{-3}P^{-2}     &  \mbox{for}\  r>r_{\rm lc},\\
   \end{cases}  
\end{equation}
where $\kappa_3=\kappa/10^{3}$ is the secondary plasma multiplicity. For extremely large multiplicity (e.g., $\kappa>10^{4}$), $r_{\rm pl}$ can be smaller than $r_{\rm cyc}$ and is estimated using a slightly different equation according to \citet{bar86}, but the difference is small and will not affect the comparison with $r_{\rm lc}$. In Fig.~\ref{fig:rad} we compare $r_{\rm pl}$ and $r_{\rm cyc}$ with $r_{\rm lc}$ as a function of spin period for $B=10^{10}$ and $10^{14}$\,G, respectively. Shaded regions in Fig.~\ref{fig:rad} show $r_{\rm pl}$ with $\gamma$ in the range of $10^2$ to $10^3$, and we note that $r_{\rm pl}$ is less sensitive to $\kappa$. For Galactic magnetars and normal pulsars, their polarisation limiting and resonance radii are much smaller than the light cylinder radius as expected. For millisecond pulsars, the polarisation limiting radius is comparable to the light cylinder radius, and swept-back field lines close to the light cylinder have been suggested to explain some polarisation features of millisecond pulsars~\citep[e.g.,][]{bar86}. More importantly, we show that for pulsars with strong magnetic fields (e.g., $>10^{14}$\,G) and short spin periods (e.g., $<0.01$\,s), $r_{\rm pl}$ and $r_{\rm cyc}$ become orders of magnitude larger than $r_{\rm lc}$. This suggests that wave mode coupling effects are negligible and will not be able to produce significant circular polarisation in the magnetosphere of fast spinning magnetars.

Unlike the wave mode coupling effect, cyclotron absorption can be strong even at $r_{\rm cyc}\gg r_{\rm lc}$ if its optical depth is large. The optical depth of cyclotron absorption, $\tau$, can be estimated as~\citep[e.g.,][]{lm01,flm03}
\begin{equation}
    \tau\approx\Gamma\epsilon r_{\rm cyc}/c,
\end{equation}
where $\Gamma$ is the absorption coefficient and $\epsilon$ is a parameter characterising the radial width of the cyclotron resonance region as $\Delta r_{\rm cyc}=\epsilon r_{\rm cyc}$. As discussed in \citet{flm03}, $\epsilon$ depends on the plasma distribution and wave propagation angle $\theta$ and is estimated to be $\sim1$ for $r_{\rm cyc}\ll r_{\rm lc}$. However, for $r_{\rm cyc}\gg r_{\rm lc}$, since the absorption coefficient drops quickly when the wave frequency deviates from the cyclotron resonance frequency, we expect $\epsilon$ to be $\ll1$. The absorption coefficient, $\Gamma$, at the cyclotron resonance frequency, $\omega_{\rm cyc}$, can be estimated as~\citep{flm03}
\begin{equation}
    \Gamma\approx\pi\frac{\omega_{\rm p}^2}{\omega_{\rm cyc}}\theta^2,
\end{equation}
where $\omega_{\rm p}$ is the plasma frequency. Adopting solutions of \citet{bar86} and using their evaluations of $\omega_{\rm p}$ and $\omega_{\rm cyc}$, we can estimate the optical depth as 
\begin{equation}
\label{eq:tau}
    \tau\approx\begin{cases}
    2700\epsilon\kappa_{3}\theta^{2}B_{12}^{1/5}\nu_{9}^{-1/5}\gamma_{2}^{-1/5}P^{2/5} &  \mbox{for}\ r\lesssim r_{\rm lc}\\
    5700\epsilon\kappa_{3}\theta^{2} &  \mbox{for}\ r>r_{\rm lc}\\
       \end{cases}  
\end{equation}
In the case of $r<r_{\rm lc}$, $\tau$ is estimated to be $\sim1$ for $\eta=100$, $\theta=0.1$ and $\epsilon\approx1$, which agrees with previous results~\citep[e.g.,][]{flm03,wlh10} and suggests marginal absorption in normal pulsar magnetosphere. For $r\gg r_{\rm lc}$, which applies to fast spinning magnetars, we find that the optical depth is mainly determined by the secondary plasma multiplicity, the size of the absorption region, and the propagation angle. Although these parameters are highly uncertain at $r\gg r_{\rm lc}$, reasonable assumptions with $\epsilon\ll1$ and $\kappa\approx100$ give us an optical depth of $\sim1$, which is necessary to explain the detection of repeating bursts. In any case, Eq.~\ref{eq:tau} suggests that cyclotron absorption could be significant if the secondary plasma density is high and the absorption region is not too small outside the light cylinder.

\begin{figure}
\centering
\includegraphics[width=0.5\textwidth]{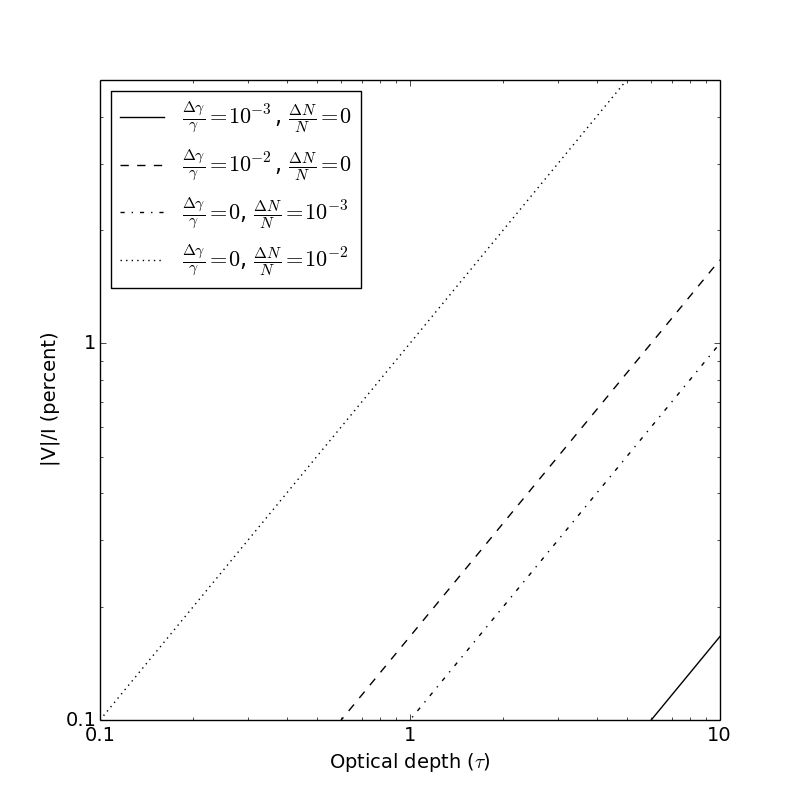}
\caption{Degrees of circular polarisation as a function of the optical depth of cyclotron absorption. Different lines show how the degree of circular polarisation changes with different levels of asymmetry in electrons and positrons.}
\label{fig:fracv}
\end{figure}

Circular polarisation can be produced by cyclotron absorption when there are asymmetries in electrons and positrons (in either their density $\Delta N/N$ or Lorentz factor $\Delta\gamma/\gamma$). As shown by \citet{wlh10}, the degree of circular polarisation produced by cyclotron absorption can be estimated as 
\begin{equation}
    \frac{V}{I}\approx\frac{e^{-\Delta\tau}-1}{e^{-\Delta\tau}+1},
\end{equation}
where $\Delta\tau$ is the difference in optical depth of electrons and positrons. $\Delta\tau$ is determined by $\Delta N/N$ and $\Delta\gamma/\gamma$ as
\begin{equation}
    \Delta\tau=2\tau\left(\frac{\Delta N}{N}-\frac{\Delta\gamma}{6\gamma}\right).    
\end{equation}
In Fig.~\ref{fig:fracv} we show the degree of circular polarisation as a function of optical depth for different $\Delta N/N$ and $\Delta\gamma/\gamma$. We can see that high degrees of circular polarisation can only be produced when both the optical depth and asymmetry in electrons and positrons are large. For marginal absorption ($\tau\sim1$) and low levels of asymmetry ($\Delta N/N\sim10^{-2}$ and $\Delta \gamma/\gamma\sim10^{-2}$), the expected circular polarisation is only a few percent.

\section{Discussion and conclusions}
\label{sec:con}

In this paper, we investigated wave mode coupling and cyclotron absorption effects in the magnetosphere of fast spinning magnetars, one of the promising origins of repeating FRBs. We show that the polarisation-limiting radius is much larger than the radius of the light cylinder of fast spinning magnetars, and therefore wave mode coupling is unlikely to produce significant circular polarisation. Depending on the secondary plasma density and the size of the absorption region, cyclotron absorption could be strong outside the light cylinder, but high degrees of circular polarisation can only be produced if the asymmetry in electrons and positrons is unusually high (e.g., $\Delta N/N\gg10^{-2}$). Our results suggest that it is challenging for propagation effects commonly discussed for radio pulsars to produce strong circular polarisation in the case of fast spinning magnetars. 

Our discussion is focused on propagation effects in the magnetosphere of fast spinning magnetars and assumes that the radio emission originates within the magnetosphere. This is one of the many FRBs models and possibly one of the many emission channels under the magnetar model~\citep[e.g.,][]{lkz20}. We did not explore the origin of circular polarisation observed in non-repeating FRBs and argue that non-repeating FRBs might involve different radiation mechanisms. We suggest that the difference in polarisation property of repeating and non-repeating FRBs, together with their diverse burst morphology, can be key discriminant of their origins and radiation mechanisms. If non-repeating FRBs are also powered by fast spinning magnetars, we argue that their high degrees of circular polarisation can not be explained by propagation effects within the magnetosphere, and other effects, such as Faraday conversion in dense and magnetised environments~\citep{gl19}, are required.

The current sample of repeating FRBs with deep polarisation observation is still limited. As more FRBs being discovered~\citep[e.g.,][]{mbb+10,scd+18,krc+19,zll+20,nll+21,aab+21}, highly sensitive polarisation observations of repeating FRBs will enable us to test models of FRB central engine and radiation mechanism. If repeating FRBs are all powered by fast spinning magnetars, we expect to see low degrees (close to zero) of circular polarisation despite deep observations. Even if circular polarisation can be detected in some repeating FRBs, the distribution of fractional circular polarisation can be compared with those of non-repeating FRBs and Galactic magnetars. A statistically lower fractional circular polarisation of repeating FRBs will provide supports to models that involve different origins and/or radiation mechanisms for repeating and non-repeating FRBs. Ultimately, if sub-second rotational periodicity is detected in repeating FRBs, propagation effects in the magnetosphere of fast spinning magnetars can be studied through variations of circular and linear polarisation across the pulse~\citep[e.g.,][]{wlh10}.

The origin and many observational properties of circular polarisation of normal radio pulsars are still not fully understood~\citep[e.g.,][]{jk18,ijw19,dwi21}. It is even harder to understand those of millisecond pulsars~\citep[e.g.,][]{dhm+15} as their polarisation-limiting regions are close to the light cylinder~\citep[e.g.,][]{jon20}. Despite the detection of strong circular polarisation from radio-loud magnetars, their origin and properties have not been investigated in the literature. For magnetars with short spin periods, effects arising in highly magnetised vacuum or highly relativistic plasma can be important~\citep[e.g.,][]{km98}. Properties of plasma and magnetic fields both inside and outside the light cylinder of magnetars are not clear, which can significantly affect our treatment of propagation effects. The asymmetry in electrons and positrons can also be much larger in pulsar magnetoshperes generating FRBs~\citep[e.g.,][]{klb+17,wang+19}, which implies that the circularisation effect can be more important than for radio pulsars. It is beyond the scope of this paper to develop a comprehensive model for the magnetosphere of fast spinning magnetars, but theoretical studies of this special case should be encouraged, especially considering its potential link with repeating FRBs. 

The detection of an extremely intense radio burst (FRB 200428) from the Galactic magnetar SGR 1935$+$2154 provides strong supports to the theory that magnetars are the origin of at least some FRBs~\citep{abb+20,brb+20}. Despite the extreme intensity and strong linear polarisation, no significant circular polarisation was detected in FRB 200428~\citep{abb+20}. A few days after the initial event, \citet{zjm+20} detected a much fainter burst from SGR~1935$+$2154 with a high degree of linear polarisation but no evidence of circular polarisation. Two more bursts from SGR 1935$+$2154 were detected in May 2020, and one of them shows some degrees of circular polarisation~\citep{ksj+21}. As a 3.2 second magnetar with a surface magnetic field strength of $B\approx2.2\times10^{14}$\,G~\citep{ier+16}, we expect propagation effects in the magnetosphere of SGR 1935$+$2154 to be significant and its polarisation properties similar to other Galactic magnetars. Further studies of polarisation properties of Galactic magnetars will provide us insights into their magnetic field structure and plasma properties. For example, \citet{hww+21} recently discussed the modification of the rotating vector model in the case of magnetars with twisted magnetic fields.

\acknowledgments
The authors thank the anonymous referee for valuable suggestions that have allowed us to improve this manuscript significantly. We thank Simon Johnston, Weiwei Zhu, Chenhui Niu, Phil Edwards, Yi Feng for discussions and comments. S. D. is the recipient of an Australian Research Council Discovery Early Career Award (DE210101738) funded by the Australian Government. W.-Y.W. acknowledge the support of NSFC grants 11633004, 11653003, CAS grants QYZDJ-SSW-SLH017, and CAS XDB 23040100, and MoST grant 2018YFE0120800, 2016YFE0100300, R.X.X. acknowledges the support of National Key R\&D Program of China (No. 2017YFA0402602), NSFC 11673002 and U1531243, and the Strategic Priority Research Program of CAS (No. XDB23010200), J.-G. L.acknowledge the support of NSFC 12003047.

\bibliography{ms}

\end{document}